\documentstyle[12pt]{article}

\begin{document}
\thispagestyle{empty}
\begin{center}
\LARGE \tt \bf{A note on the Limits to Global Rotation from Teleparallel G\"{o}del Universe}
\end{center}
\vspace{1cm}
\begin{center} {\large L.C. Garcia de Andrade\footnote{Departamento de Fisica Teorica-IF-UERJ-CEP:20550-013,Rio de Janeiro,RJ}}
\end{center}
\vspace{1.0cm}
\begin{abstract}
A teleparallel G\"{o}del universe is shown to lead to a simple and natural relation between Cartan torsion and the global rotation of the universe.This is straightforward if one uses the formalism of Cartan's calculus of exterior differential forms.It is possible to place a limit on the global rotation from the knowledge of torsion and the converse is also true.The Obukhov recent limit of $\frac{\omega}{H_{0}}=1$ leads to the well know value for cosmological torsion of $10^{-17} cm^{-1}$.COBE constraints on the temperature anisotropy allows us to obtain limits on the torsion fluctuation in G\"{o}del's universe. 
\end{abstract}      
\vspace{1.0cm}       
\begin{center}
\Large{PACS number(s) : 0420,0450}
\end{center}
\newpage
\pagestyle{myheadings}
\markright{\underline{Global rotation and G\"{o}del Teleparallel cosmology}}
\paragraph*{}
Limits on the global rotation of the universe have been discussed on the realm of Bianchi anisotropic cosmologies \cite{1} in the context of General Relativity (GR).Bunn et al \cite{2} for example placed limits on shear and global rotation of the ratio between the global rotation to the present universe value of Hubble constant $H_{0}$ as $\frac{\omega}{H_{0}}<10^{-6}$.More recently \cite{3} has shown by making Einstein-Cartan (EC) cosmology that this limit could reach values as big as  as $\frac{\omega}{H_{0}}=1$!In this short note we show that a very simple and straightforward way to achieve a relation between Cartan's torsion \cite{4} Q and the global rotation ${\omega}$ \cite{5} is through the use of a G\"{o}del Teleparallel model \cite{6} with help of Cartan's calculus of exterior differential forms.G\"{o}del metric \cite{1} is given wrt the one-form basis ${\omega}^{i}$ $(i=0,1,2,3)$ by the expression
\begin{equation} 
ds^{2}=-({\omega}^{0})^{2}+({\omega}^{1})^{2}+({\omega}^{2})^{2}+({\omega}^{3})^{2}
\label{1}
\end{equation}
where
\begin{equation}
{\omega}^{0}=\frac{1}{\sqrt{2}{\omega}}(dt+e^{x}dz)
\label{2}
\end{equation}
\begin{equation}
{\omega}^{1}=\frac{1}{\sqrt{2}{\omega}}dx
\label{3}
\end{equation}
\begin{equation}
{\omega}^{2}=\frac{1}{\sqrt{2}{\omega}}dy
\label{4}
\end{equation}
\begin{equation}
{\omega}^{0}=\frac{1}{\sqrt{2}{\omega}}(dt+e^{x}dz)
\label{5}
\end{equation}
From the Cartan's structure equations
\begin{equation}
Q^{i}=d{\omega}^{i}+{\omega}^{i}_{k}{\wedge}{\omega}^{k}
\label{6}
\end{equation}
\begin{equation}
R^{i}_{j}=d{\omega}^{i}_{j}+{\omega}^{i}_{k}{\wedge}{\omega}^{k}_{j}
\label{7}
\end{equation}
where 
\begin{equation}
R^{i}_{j}({\Gamma})=\frac{1}{2}R^{i}_{jkl}{\omega}^{k}{\wedge}{\omega}^{l}
\label{8}
\end{equation}
is the curvature 2-form and $R^{i}_{jkl}$ is the Riemann-Cartan (RC) geometry curvature tensor $U_{4}$ geometry ,where ${\Gamma}$ is the RC connection of the manifold.One immediatly notices that a minimal teleparallel geometry is obtained by making use of the constraint \cite{7,8,9}
\begin{equation}
{\omega}^{i}_{j}=0
\label{9}
\end{equation}
since from equation (\ref{7}) this condition implies teleparallel condition of vanishing RC curvature 
\begin{equation}
R^{i}_{jkl}({\Gamma})=0
\label{10}
\end{equation}
However from expression (\ref{6}) one notices as well that torsion is not trivial and is given by
\begin{equation} 
Q^{i}=d{\omega}^{i}
\label{11}
\end{equation}
From (\ref{11}) and G\"{o}del metric in terms of the one-form basis ${\omega}^{i}$ one obtains the following components of torsion forms 
\begin{equation}
Q^{0}=d{\omega}^{0}=2{\omega}{\omega}^{1}{\wedge}{\omega}^{3}
\label{12}
\end{equation}
\begin{equation}
Q^{1}=d{\omega}^{1}=0
\label{13}
\end{equation}
\begin{equation}
Q^{2}=d{\omega}^{2}=0 
\label{14}
\end{equation}
\begin{equation}
Q^{3}=d{\omega}^{3}={\sqrt{2}}{\omega}{\omega}^{1}{\wedge}{\omega}^{3}
\label{15}
\end{equation}
From these last four expressions and the definition of torsion forms $Q^{i}$ in terms of torsion tensor $T^{i}_{jk}$
\begin{equation}
Q^{i}=\frac{1}{2}T^{i}_{jk}{\omega}^{j}{\wedge}{\omega}^{k}
\label{16}
\end{equation}
one obtains\begin{equation}
T^{0}_{13}=4{\omega}
\label{17}
\end{equation}
\begin{equation}
T^{3}_{13}=2\sqrt{2}{\omega}
\label{18}
\end{equation}
Since torsion is then proportional to the global rotation ${\omega}$ and $H_{0}=10^{-17}cm^{-1}$ one may conclude that if one takes $\frac{\omega}{H_{0}}=1$ one obtains
\begin{equation}
Q=H_{0}=10^{-17}cm^{-1}
\label{19}
\end{equation}
This value of torsion has been obtained by L\"{a}mmerzahl \cite{10} by investigating the splitting of the spectra of atoms in the presence of a torsion field on terestrial laboratory.On the other hand if one assumes that the torsion of the Universe given by de Sabbata and Sivaram \cite{11} and L\"{a}mmerzahl \cite{10} as $Q=10^{-17}cm^{-1}$ and the expressions (\ref{12}) and (\ref{18}) one obtains that a possible limit on global rotation is given by $\frac{\omega}{H_{0}}=1$ exactly confirming Obukhov's result.Recently other teleparallel solutions like Kerr and Schwarzschild metrics.A simple reasoning which follows shows that it is possible to place limits on the torsion fluctuations over the constant global rotation of the universe.Since 
\begin{equation}
{\omega}{\alpha}Q
\label{20}
\end{equation}
and the relation between rotation and the matter density is ${\omega}^{2}=8{\pi}{\rho}$ from this expression we obtain
\begin{equation}
{\omega}{\alpha}Q
\label{21}
\end{equation}
and 
\begin{equation}
\frac{{\delta}{\omega}}{{\omega}_{0}}=\frac{1}{2}\frac{{\delta}{\rho}}{\rho}=\frac{1}{2}\frac{{\delta}T}{T}
\label{21}
\end{equation}
Since $Q_{0}=10^{-17}cm^{-1}$ and by COBE constraint $\frac{{\delta}T}{T}=10^{-5}$ one obtains for the torsion fluctuation around the G\"{o}del Teleparallel $T_{4}$ model ${\delta}Q=10^{-22}cm^{-1}$.Finally we would like to mention that earlier M\"{u}ller-Hoissen and J.Nitsch \cite{12} have pointed out that G\"{o}del model in Teleparallel gravity would not be a viable model for the Universe.Of course this has in a certain sense be minimized here since as we show the G\"{o}del teleparallel cosmology is found to be a model which leads naturally to a measure of global rotation of the universe in terms of torsion.
\section*{Acknowledgments}
\paragraph*{}
Thanks are due to Prof.Yu N. Obukhov and Prof.Rudnei Ramos, for their constant advice on the subject of this paper.I am very much indebt to CNPq. (Brazilian Government Agency) for financial support.


\begin{thebibliography}{7}
\bibitem{1}I.Ciufolini and J.A.Wheeler,Gravitation and Inertia,(1995),Princeton University Press.
\bibitem{2}E.Bunn,P.Ferreira andJ.Silk,How Anisotropy is our Universe?,astro-ph/9605123.
\bibitem{3}Yu N.Obukhov,Physical foundations and observations of the rotation problem in Cosmology,astro-ph/0008106.
\bibitem{4}E.Cartan and A.Einstein,Letters on Absolute Parallelism,(1979),Princeton University Press.
\bibitem{5}V.de Sabbata and M.Gasperini,Introduction to Gravitation,(1985)World Scientific. 
\bibitem{6}Garcia de Andrade,Found. of Physics,(1990).
\bibitem{7}J. Nester,Class. and Quantum Gravity,(1988).
\bibitem{8}W.Kopczynski,J.Phys.A,(1982).
\bibitem{9}P.S.Letelier,Class. and Quantum Gravity 12 (1995),2221.
\bibitem{10}C.L\"{a}mmerzahl,Phys.Lett.A,(1997).
\bibitem{11}V.de Sabbata and C.Sivaram,Spin and Torsion in Gravitation,(1994)World Scientific. 
\bibitem{12}M.Hoissen and J.Nitsch,Phys.Rev.D 28,(1983),718 and Gen.Rel. and Gravitation (1982),8,747.
\end{thebibliography}
\end{document}